\documentclass[12pt]{article}

\newcommand{\beq}[1]{\begin{equation}\label{#1}}
\newcommand{\eeq}{\end{equation}}
\newcommand{\bear}[1]{\begin{eqnarray}\label{#1}}
\newcommand{\ear}{\end{eqnarray}}

\newcommand{\tr}{ {\rm tr} }
\newcommand{\GUT}{ {\rm GUT} }
\newcommand{\const}{ {\rm const} }

\newcommand{\fnm}{\footnotemark}
\newcommand{\fnt}{\footnotetext}

\begin{document}

 \begin{center}
 \large \bf
  On Time Variations of  Gravitational  and Yang-Mills Constants in a Cosmological Model of Superstring
  Origin

 \end{center}

 \vspace{0.3truecm}

 \begin{center}

 \normalsize\bf

  V. D. Ivashchuk \fnm[1]\fnt[1]{e-mail: ivashchuk@mail.ru}
  and V. N. Melnikov \fnm[2]\fnt[2]{e-mail:  melnikov@phys.msu.ru},

 \vspace{0.3truecm}

 \it Center for Gravitation and Fundamental Metrology, VNIIMS,
 Ozyornaya 46, Moscow 119361, Russia and

 \it Institute of Gravitation and Cosmology, Peoples' Friendship
 University of Russia,  Miklukho-Maklaya 6, Moscow 117198,
 Russia

 \end{center}

\begin{abstract}

In the framework of $10$-dimensional ``Friedmann-Calabi-Yau''
cosmology of superstring origin we show that the time variation of
either Newton's gravitational constant or Yang-Mills one is
unavoidable  in the present epoch.

\end{abstract}

\section{Introduction}

The idea of time variations of Newton's gravitational constant
originally proposed by Dirac  \cite{Dir} (for a review see
\cite{MelSt}) acquired more importance with the appearance of
superstring theories \cite{SchGr,GHMR}.

The present  experimental constraints on time variations of the
gravitational constant  have the form

\begin{equation} \label{0.2}
        |\dot{G}/(GH)| < 0.001,
\end{equation}
 where  $H$  is the Hubble parameter, see \cite{P} and references therein.

Here we consider the ``Friedman-Calabi-Yau'' (FCY) cosmology based
on the \\
 ten-dimensional $SO(32)$ or $E_8\times E_8$ Yang-Mills
supergravity theory \cite{Ch} with a Lorentz Chern-Simons
three-form, introduced by Green and Schwarz \cite{SchGr} for
anomaly cancellation, and with the Gauss-Bonnet term, introduced
in \cite{Zw,CHSW}. These additional terms are of  superstring
origin \cite{GW}: they appear as the next to leading terms in the
$\alpha '$ decomposition ($\alpha '$ is the string parameter) of
the Fradkin-Tseytlin effective action \cite{FTs} for a heterotic
string \cite{GHMR} (see \cite{MTs}). The supergravity action is a
leading term in this decomposition.

In the open-universe case of FCY cosmology with dust matter $\rho
>0$, $p_3=p_6=0$ ($\rho >0$  is the density and $p_3,p_6$ are
pressures, see (\ref{9}) below) and a constant dilaton field,
$\varphi (t)= \const$, Wu and Wang calculated the present value of
${\dot{G}/G}$ \cite{WuW} and obtained the estimate

\begin{equation}
(\dot{G}/G)_0 \approx -10^{-11 \pm 1} \ {\rm yr}^{-1}
\label{0.1} .
\end{equation}

This asymptotic relation was explained and generalized in
\cite{BIM}, by simply using  the equations of motion for a
multidimensional cosmological model with anisotropic fluid, which
describe the asymptotical behavior of FCY model with $\varphi=
\const$.

 Here we present a corrected and updated version of our
 old paper \cite{IM-88}, which was devoted to the problem of
 variation of  $G$ in FCY cosmological model with the restrictions
 on the density $\rho >0$ and $3$-dimensional pressure $p_3 \geq 0$ imposed \cite{WuW}.
 At that times such restrictions looked  physical.
Now, we consider the problem of  simultaneous stability of the
effective gravitational constant $G$ and the effective Yang-Mills
constant $g_{\GUT}$ of Grand Unification Theory (GUT),  which
arise in this $10D$ model. In what follows we do not put any
restrictions  on $\rho$,  $p_3$ and $p_6$. We prove that in the
FCY cosmology there are no solutions to equations of motion with a
constant radius of  internal space and a constant dilation field,
which do not contradict the present accelerated expansion of our
Universe \cite{Riess,Perl}. Hence in FCY cosmology a simultaneous
stability of $G$ and $g_{\GUT}$ is  impossible in the present
epoch.

\section{The model}

 We take the action of the model as in \cite{WuW}

\begin{eqnarray}
S = \int d^{10}x\sqrt{-g} \left\{ \frac{1}{2\kappa^2}R -
 \frac{3}{4}\kappa^2\varphi^{-3/2}H^2_{MNP}
 - \frac{9}{16\kappa^2}(\varphi^{-1}\partial_M
 \varphi )^2 \right. \nonumber    \\ \left.
 -\frac{1}{4}\varphi^{-3/4} \left[ \frac{1}{30} \tr F^2_{MN} +
 (R^2_{MNPQ}-4R^2_{MN}+ R^2) \right] \right\} + S_F\ \label{1} ,
\end{eqnarray}
where $g_{MN}$ and $\varphi$ are the metric and dilation fields,
$F_{MN}$ and $H_{MNP}$ are the Yang-Mills and Kolb-Ramond field
strengths

 $$ F = \frac{1}{2} F_{MN}dx^M\wedge dx^N = dA + A \wedge A, $$
where $A=A_Mdx^M$ is the one-form with the value in the Lie
algebra ad $\hat{g}$, $\hat{g} = so(32)$, $e_8\oplus e_8$ (${\rm
ad } \hat{g}$ is the image of the adjoint representation of
$\hat{g}$,  ${\rm ad} \hat{g} \approx \hat{g}$ for any semi-simple
Lie algebra $\hat{g}$);

\begin{equation}
H = \frac{1}{3!}H_{MNP}dx^M\wedge dx^N\wedge dx^P = dB -
\omega_{3Y} + \omega_{3L}\ \label{2},
\end{equation}
where $B=\frac{1}{2}B_{MN}dx^M\wedge dx^N$ is a two-form,
$\omega_{3Y}$ is the Yang-Mills Chern-Simons three-form,

\begin{equation}
 \omega_{3Y} = \frac{1}{30} {\rm tr}(A\wedge F-\frac{1}{3}A\wedge A\wedge
 A),  \label{3a}
\end{equation}
and $\omega_{3L}$ is the Lorentz Chern-Simons three-form,

\begin{equation}
 \omega_{3L}= {\rm tr}(\omega \wedge \Omega -\frac{1}{3}\omega \wedge
 \omega \wedge \omega ). \label{3b}
\end{equation}

In (\ref{3b}) $\omega =\omega_Mdx^M$ is the spin connection, which
is a one-form with the value in $so(1,9)$:

$$ \omega_M = \mid \mid \omega^A_{BM}\mid \mid = \mid \mid e^A_N
 \nabla_Me^N_B\mid \mid \subset so(1,9), $$

 $e^A_N$ is the basis
(zehnbein) of vector fields which diagonalizes the metric

 $$g_{MN}  = e^A_M e^B_N \eta_{AB}, $$

 $\eta_{AB} = {\rm diag}(-1,\ +1,\ ...,\ +1)$,  $\Omega$ is the curvature two-form:

 $$ \Omega = d\omega + \omega \wedge \omega. $$

 In (\ref{1}) $\kappa^2$ is the $10$-dimensional gravitational constant
 and   $S_F$ is the Fermi part of the action \cite{Ch}. Here we
 deal with pure bosonic solutions to equations of motion (i.e., with zero Fermi fields) and
 hence an explicit relation for $S_F$ is irrelevant for our consideration.

The action (\ref{1}) and the energy-momentum tensor $T_{MN}$ lead
to the following equations of motions (see \cite{WuW} for $\varphi
= \const$):
\begin{eqnarray}
R_{MN}-\frac{1}{2}g_{MN}R = \frac{9}{2}\kappa^4 \varphi^{-3/2}
\biggl( H_{MPQ}H_N^{\ \ \ PQ}-\frac{1}{6}g_{MN}H^2_{PQS} \biggr)
    \nonumber \\
 + 9\kappa^4\nabla^S(\varphi^{-3/2}H_{MPQ}R_{SN}^{\ \ \ \ PQ}) +
 \frac{9}{8}\varphi^{-2}  \biggl[ \partial_M \varphi \partial_N\varphi
 -\frac{1}{2}g_{MN}(\partial_P\varphi )^2 \biggr]
    \nonumber \\
 + \frac{1}{30}\kappa^2\varphi^{-3/4} \biggl( \tr
 F_{MP}F_N^{\ \ P}-\frac{1}{4}g_{MN}\ \tr F^2_{PQ} \biggr)
        \qquad  \qquad     \nonumber \\
 -\frac{1}{2}\kappa^2
 \varphi^{-3/4} \biggl[ \frac{1}{2}g_{MN}(R^2_{PQST}-4R^2_{PQ}+R^2)
 - 2RR_{MN}      \nonumber \\
 + 4R_{MP}R_N^{\ \ P} + 4R_{MPNQ}R^{PQ} - 2R_M^{\ \ PQS}R_{NPQS} \biggr] + \kappa^2
 T_{MN} + D_{MN} ,  \label{4}
\end{eqnarray}
\begin{eqnarray}
 \nabla_M(\varphi^{-3/2}H^{MNP}) = 0,   \label{5}\\
 D_M(\varphi^{3/4}F^{MP}) + 9\kappa^2(\varphi^{-3/2}F_{MN}H^{MNP})
 = 0, \label{6} \\
 6\nabla_M(\varphi^{-2}\partial^M\varphi) +
 6\varphi^{-3}(\partial_M\varphi)^2 +
 6\kappa^4\varphi^{-5/2}H^2_{MNP} \qquad \nonumber \\
  \qquad \qquad \qquad \qquad + \kappa^2\varphi^{-7/4}\left[\frac{1}{30}
 \tr F^2_{MN}+(R^2_{MNPQ}-4R^2_{MN}+R^2) \right] = 0\ .  \label{7}
\end{eqnarray}
In (\ref{4}) $D_{MN}$ is a term with  derivatives of the dilaton
field  which appears from the variation of the Gauss-Bonnet term.
This $D$-term vanishes for $\varphi = \const$.

Let us consider the ten-dimensional manifold
\begin{equation}
   M^{10} = R \times M^3_k \times K ,    \label{8}
\end{equation}
where $M^3_k=S^3,~R^3,~L^3$ for $k=+1,0,-1$, respectively, and $K$
is a Calaby-Yau manifold, i.e., a compact 6-dimensional K\"ahler
Ricci-flat manifold with the $SU(3)$ holonomy group.

Let the energy-momentum tensor be
\begin{equation}
  T = T_{MN}dx^M\otimes dx^N = \rho (t)dt\otimes
  dt + p_3(t)a^2_3(t)g^{(3)} + p_6(t)a^2_6(t)g^{(6)}\ ,
 \label{9}
\end{equation}
 where $g^{(3)}$ and $g^{(6)}$ are metrics on $M^3_k$ and $K$, $\rho (t)$ is
 the  energy density, $p_3(t)$ and $p_6(t)$ are pressures corresponding
to $M^3_k$ and $K$.

The system (\ref{4})-(\ref{7}) on the manifold (\ref{8}) with the
source (\ref{9}) and the following ansatz
\begin{eqnarray}
   g^{(10)} = -dt\otimes dt + a^2_3(t)g^{(3)} + a^2_6(t)g^{(6)},
   \label{10} \\
  H = 0, \label{11} \\
  \varphi = \varphi (t), \label{12} \\
  A = {\rm ad}(i(\omega^{(6)})).
 \label{13}
\end{eqnarray}
leads to a cosmology model, which we call the
``Friedman-Calaby-Yau'' (FCY) cosmology. In (\ref{13})
$\omega^{(6)}$ is the spin connection on $K$ corresponding to the
basis $e^{(6)\alpha}$, which diagonalizes $g^{(6)}$; $i :
so(6)\rightarrow \hat{g}$ is the enclosure of the Lie algebra
$so(6)$ (in the case $\hat{g} = e_8\oplus e_8$, $i$ may be
defined, for example, with the aid of the decomposition
\cite{GHMR}: $e_8=so(16)\oplus V_{128}$). It follows from
(\ref{13}) that

\begin{equation}
 F = {\rm ad}(i(\Omega^{(6)})),  \label{13a}
\end{equation}
where $\Omega^{(6)}= d \omega^{(6)} + \omega^{(6)}\wedge
\omega^{(6)}$.

From (\ref{13}) and the trace identity
\begin{equation}
  \frac{1}{30} {\rm tr} ({\rm ad}(i(X)) {\rm ad}(i(Y))) = {\rm tr}(XY)
   \label{14}
\end{equation}
for all $X, Y \in so(6)$, we have
\begin{equation}
 \omega_{3Y} = \frac{1}{30} {\rm tr} \biggl( \frac{2}{3}A\wedge
 F+\frac{1}{3}A\wedge dA \biggr) = {\rm tr} \biggl( \frac{2}{3}\omega^{(6)}\wedge
 \Omega^{(6)} + \frac{1}{3}\omega^{(6)}\wedge d\omega^{(6)} \biggr) =
 \omega^{(6)}_{3L}. \label{15}
\end{equation}

 It can be verified that in the basis $(e^{(10)A})=(dt,\ a_3(t)e^{(3)a},\ a_6(t)e^{(6) \alpha})$,
where $e^{(3)a}$ is the basis on $M^3_k$ diagonalizing $g^{(3)}$,
we obtain
\begin{equation}
 \omega_{3L} = \omega^{(3)}_{3L} + \omega^{(6)}_{3L} + f_3, \qquad
 \hspace{6cm}  \label{16}
\end{equation}
where $df_3=0$ and

$$\omega^{(3)}_{3L}= {\rm tr}(\omega^{(3)}\wedge \Omega^{(3)} -
\frac{1}{3}\omega ^{(3)}\wedge \omega^{(3)}\wedge \omega^{(3)}),$$
$\omega^{(3)}$ is the spin connection on $M^3_k$ corresponding to
$e^{(3)a}$. From (\ref{2}), (\ref{15}) and (\ref{16}) we have

\begin{equation}
H = dB + \omega^{(3)}_{3L} + f_3.  \label{17}
\end{equation}

It follows from (\ref{17}) and $d\omega^{(3)}_{3L}=df_3=0$ that
for every domain $\Omega \subset M^{10}$ with $H^3(\Omega ,R)=0$
there is some $B$ such that $H=0$.

The spin connection $\omega^{(6)}$ on $K$ obeys the identity
\begin{equation}
 D_m(\omega^{(6)})\Omega^{(6)mn} = 0, \label{18}
\end{equation}
where $D_m(\omega)=\nabla_m + [\omega_{m},.]$. The identity
(\ref{18}) is equivalent to
 $$\nabla^{(6)}_m R^{(6)mnpq} = 0$$
and is valid for any K\"ahler Ricci-flat manifold \cite{WitW}.
Equation (\ref{6}) is satisfied identically due to (\ref{18}),
(\ref{10})-(\ref{13}) and (\ref{13a})
$(D_M=D_M(A)=\nabla_M+[A_M,.])$; (\ref{5}) is satisfied owing to
(\ref{11}).

 Now we put
\begin{equation}
a_6(t)= {\rm const}, \qquad   \varphi(t) = {\rm const}. \label{27}
\end{equation}

What is the reason for the  restrictions (\ref{27})? For the
cosmology under consideration the effective $4D$ gravitational
constant reads

\begin{equation}
 G = \const \cdot a^{-6}_6(t).               \label{29}
\end{equation}
while the  effective $4D$ Yang-Mills constant has the following
form
\begin{equation}
 g_{\GUT} = {\rm const} \cdot a^{-3}_6(t) \varphi^{3/8}(t).  \label{30}
\end{equation}

This  follows  from the action (\ref{1}) and the ansatz (\ref{10})
for the metric. It is obvious that the stability of $G$ and
$g_{\GUT}$ is equivalent to the condition (\ref{27}). Thus we are
interested in solutions with stable effective constants $G$ and
$g_{\GUT}$.

Then, equations (\ref{4}) and (\ref{7}) in the ansatz
(\ref{10})-(\ref{13}) may be rewritten as follows \cite{IM-88}:
\begin{eqnarray}
 3a_3^{-2}(k + \dot{a}^2_3) =
  \kappa^2\rho,  \label{4a} \\
 a_3^{-2}(k + \dot{a}^2_3+ 2a_3 \ddot{a}_3) =
  - \kappa^2 p_3,  \label{4b}  \\
 a_3^{-2}(k+ \dot{a}^2_3 + a_3 \ddot{a}_3) =
  - (1/3)p_6  \nonumber \\
 + 2\kappa^2 \varphi^{-3/4}a_3^{-3}\ddot{a}_3 (k + \dot{a}^2_3),
       \label{4c}  \\
 4\kappa^2\varphi^{1/4}a^{-3}_3\ddot{a}_3 (k + \dot{a}^2_3)
  = 0 . \label{7a}
\end{eqnarray}

  Here $24 a_3^{-3}\ddot{a}_3 (k +
 \dot{a}^2_3)= GB^{(4)}$   is the Gauss-Bonnet term
 corresponding to the \\
  4-dimensional part of the metric.

Equations (\ref{4a})-(\ref{4c}) and (\ref{7a}) are obtained from
(\ref{4}) and (\ref{7}), respectively, using the Ricci flatness of
$K$ and the equality

$$R^{(6)}_{pqmn} R^{(6)qp'mn} =   \frac{1}{30} {\rm tr} \
F_{mn}F_{pq}g^{(6)mp'}g^{(6)nq},$$
which follows from (\ref{13a}),
(\ref{14}) and the relation

$$ R^{(6)p}_{\ \ \ \ qmn} = e^{(6)p}_{\alpha}
e^{(6) \beta}_q \Omega^{(6)\alpha}_{\ \ \ \ \beta mn} \ .$$

Due to (\ref{4})-(\ref{7}) we get

\begin{equation}
 \nabla_{M}T^{MN} = 0.  \label{19}
\end{equation}

Relation (\ref{19}) with the substitution of (\ref{9}), (\ref{10})
and (\ref{27}) is equivalent to

\begin{equation}
  \dot{\rho} +
 3a^{-1}_3 \dot{a}_3(\rho + p_3)  = 0.  \label{19a}
 \end{equation}

\section{Solutions with constant $a_6$ and $\varphi$. }

It follows from (\ref{7a}) that

\begin{equation}
 \ddot{a}_3(k+ \dot{a}^2_3) = 0. \label{22}
\end{equation}

({\bf A.}) Consider the case  $\ddot{a}_3 \neq 0$. Then we
          get from (\ref{22})
 \begin{equation}
  k + \dot{a}^2_3 = 0 \label{23}
 \end{equation}
 which implies either $\dot{a}_3 = 0$ for $k = 0$ or $\dot{a}_3 = \pm
 1$ for $k = -1$. In both cases we get $\ddot{a}_3 = 0$,
 which  contradicts  our suggestion $\ddot{a}_3 \neq
 0$.

 ({\bf B.}) Now we put $\ddot{a}_3 = 0$. Then we obtain $\dot{a}_3 =
 C$ and $a_3 = Ct +a_0$, where $C$ and $a_0$ are constants.
 First we consider the case (\ref{23})
  which implies either $C = 0$ for $k = 0$ or $C = \pm
 1$ for $k = -1$. For the density and pressures we have
  \begin{equation}
  \rho = 0, \qquad p_3 = 0, \qquad p_6 = 0.  \label{23a}
  \end{equation}

For $k = 0$,  $C = 0$,  we get a static configuration \cite{CHSW}.
For $k = -1$, we obtain $C = \pm 1$,  the 4-dimensional part of
the metric is flat. It is the well-known Milne solution which is
isomorphic to the upper light cone (for $t > t_0$, $a_0 = - C
t_0$) in the Minkowski space. Thus we also obtain a part of a
static configuration in this case.

Consider the case $k + C^2 \neq 0$. We obtain
 \begin{equation}
  \rho \neq 0, \qquad p_3 = - \rho/3, \qquad p_6 = - \rho.  \label{24}
  \end{equation}

From (\ref{4a}) we get
\begin{equation}
  \rho = \rho_0 a^{-2}_3,   \label{25}
  \end{equation}
where $\kappa^2  \rho_0 = 3 (k + C^2) \neq 0$.

Thus, in the FCY cosmology there are no solutions to the equations
of motion obeying (\ref{27}), or equivalently, with stable
constants $G$ and $g_{\GUT}$, which are compatible with the modern
cosmological data \cite{Riess,Perl}
\begin{equation}
  \dot{a}_3 > 0, \qquad  \ddot{a}_3 > 0.   \label{28}
\end{equation}

\section{Conclusions}

We have shown that in the framework of $10D$
``Friedmann-Calabi-Yau'' cosmology of superstring origin  the
simultaneous stability of Newton's gravitational constant $G$ and
Yang-Mills constant $g_{GUT}$ is impossible  in the present epoch.
Thus one of the constants should be varying.

We note that the  Yang-Mills running constant $g_{\GUT}(\mu)$
($\mu^2 = q^2$, where $q$ is $4D$ momentum) is used for generating
the running constants for electromagnetic, weak and strong
interactions, i.e., $\alpha(\mu)$, $\alpha_w(\mu)$
$\alpha_s(\mu)$, respectively \cite{M,IM-izmt}. The temporal
stability of $g_{\GUT}(\mu)$ implies the temporal stability of the
constants $\alpha(\mu)$, $\alpha_w(\mu)$ and $\alpha_s(\mu)$ for
fixed $\mu$.

An open problem here is to find exact solutions for a FCY
cosmological model with small enough variations of $G$ and
$\alpha$ obeying modern observational restrictions.
(Multidimensional cosmological models with small enough variations
of $G$ were considered  previously in \cite{DIKM,AIKM,IKMN,Gol}
while variations of $\alpha$ and $G$ were obtained recently in
\cite{BS,BMRS,BKM} in the framework of nonlinear multidimensional
gravity.) This can be a subject of a separate study.


\begin{thebibliography}{9}

 \bibitem{Dir}
 P.A.M. Dirac,
 {\it Nature (London) } {\bf 139}, 323 (1937).

 \bibitem{MelSt}
 V.N. Melnikov and K.P. Stanjukovich, Hydrodinamics, fields and
 constants in the theory of gravitation (Energoatoizdat, Moscow,
 1983) (in Russian).

 \bibitem{SchGr}
  J.H. Schwarz, {\it Phys. Rep. } {\bf 89}, 223 (1982); \\
  M.B. Green, {\it Surv. High Energy Phys. } {\bf 3}, 127 (1983); \\
  M.B. Green and  J.H. Schwarz, {\it Phys. Lett. } {\bf B 149}, 117
  (1984).

 \bibitem{GHMR}
  D.J. Gross, J.A. Harvey, E. Martinec and R. Rohm,
  {\it Phys. Rev. Lett. } {\bf  54}, 502 (1984);
  {\it Nucl. Phys. } {\bf B 256}, 253 (1986).

  \bibitem{P}
   E.V. Piteva, {\it Astron. Lett.}  {\bf 31},  340 (2005). \\
   E.V. Pit'eva, In: Proc. of the Workshop on Precision Physics and
   Fundamental Physical Constants, December 1-4, 2009, JINR, Dubna,
   Russia, p. 53.

   \bibitem{Ch}
  A.H. Chamseddine,  {\it Nucl. Phys. } {\bf B  185}, 403 (1981);
  \\
  E. Bergshoeff, M. de Roo, B. de Witt and P. Nieuwenhuizen,
  {\it Nucl. Phys. } {\bf B 15}, 97 (1982); \\
  G.F. Chapline and N.S. Manton, {\it Phys. Lett. } {\bf B  120}, 105
  (1983).

  \bibitem{Zw}
  B. Zwiebach, {\it Phys. Lett. } {\bf B 156}, 315
  (1985).

   \bibitem{CHSW}
   P. Candelas, G.T. Horowitz, A. Strominger and E. Witten,
   {\it Nucl. Phys. } {\bf B 256}, 46 (1985).

   \bibitem{GW}
   D.J. Gross and E. Witten,
   {\it Nucl. Phys. } {\bf B  277}, 1 (1986).

   \bibitem{FTs}
   E.S. Fradkin and A.A. Tseytlin,
   {\it  Phys. Lett. } {\bf B  158}, 316 (1985);
   {\it Nucl. Phys. } {\bf B 261}, 1 (1986).

    \bibitem{MTs}
   R.R. Metsaev and A.A. Tseytlin,
   {\it  Phys. Lett. } {\bf B 185}, 52 (1987).

   \bibitem{WuW}
   Y.-S. Wu and Z. Wang,
   {\it  Phys. Rev. Lett. } {\bf  B 57}, 1978 (1986).

  \bibitem{BIM}
   K.A. Bronnikov, V.D. Ivashchuk and V.N. Melnikov,
    {\it Nuovo  Cimento}  {\bf B 102},  209-215 (1988).

  \bibitem{IM-88}
    V.D. Ivashchuk  and  V.N. Melnikov,
    {\it Nuovo  Cimento} {\bf B 102}, 131-138 (1988).

  \bibitem{Riess}
    A.G. Riess {\it et al}, {\it AJ} {\bf 116}, 1009 (1998).

    \bibitem{Perl}
    S. Perlmutter {\it et al}, {\it ApJ} {\bf 517}, 565 (1999).

    \bibitem{WitW}
    L. Witten and E. Witten,
   {\it Nucl. Phys. } {\bf B  281}, 109 (1987).

    \bibitem{M}
    W.J. Marciano, {\it Phys. Rev. Lett.} {\bf 52}, 489 (1984).

    \bibitem{IM-izmt}
    V.D. Ivashchuk and V.N. Melnikov, {\it Izmerit. Tekhnika} {\bf 7},  3-6 (1986) (in
    Russian).

    \bibitem{DIKM}
    H. Dehnen, V.D. Ivashchuk, S.A. Kononogov and V.N. Melnikov,
    {\it Grav.  Cosmol.} {\bf 11}, No. 4, 340-344 (2005); gr-qc/0602108.

    \bibitem{AIKM}
     J.-M. Alimi, V.D. Ivashchuk, S.A. Kononogov and V.N. Melnikov,
    {\it Grav.  Cosmol.} {\bf 12}, No. 2-3,
    173-178 (2006); gr-qc/0611015.

   \bibitem{IKMN}
    V.D. Ivashchuk, S.A. Kononogov, V.N. Melnikov and M.
    Novello, {\it Grav.  Cosmol.} {\bf 12},  No. 4,
    273-278 (2006); hep-th/0610167.

    \bibitem{Gol}
    A.A. Golubtsova,
    {\it Grav.  Cosmol.} {\bf 16},  No. 4,
    298-306 (2010).

    \bibitem{BS}
    K.A. Bronnikov and M.V. Skvortsova,
    {\it Grav.  Cosmol.} {\bf 19},  No. 2, 114-123
    (2013).

    \bibitem{BMRS}
    K.A. Bronnikov, V.N. Melnikov, S.G. Rubin and I.V. Svadkovsky,
    {\it Gen. Relat. and Grav.}  {\bf 45}, No. 12,
    2509-2528 (2013).

    \bibitem{BKM}
    K.A. Bronnikov, S.A. Kononogov and V.N. Mel'nikov,
     {\it Measur. Techn.} {\bf 56}, No 1, 8-16 (2013).



 \end{thebibliography}
 \end{document}